\documentclass{webofc}
\usepackage[varg]{txfonts}   \usepackage{amssymb,amsmath}
\usepackage{graphicx}
\usepackage{hyperref}
\usepackage{tikz}
\usetikzlibrary{shapes, arrows}
\usetikzlibrary{fit, calc}
\usetikzlibrary{positioning}

\newcommand{\osquared}{\ensuremath{\mathrm{O}^{2}}}

\newcommand{\secRef}[1]{Sec.~\ref{#1}}

\newcommand{\figRef}[1]{Fig.~\ref{#1}}
\newcommand{\figureRef}[1]{Figure~\ref{#1}}

\begin{document}
\title{ALICE Overwatch: Online monitoring and data quality assurance using HLT data}

\author{\firstname{Raymond} \lastname{Ehlers}\inst{1}\fnsep\thanks{Corresponding author \email{raymond.ehlers@yale.edu}} \and
        \firstname{James} \lastname{Mulligan}\inst{1}
}

\institute{Department of Physics, Yale University, New Haven, CT, USA}

\abstract{ALICE Overwatch is a project started in late 2015 to provide augmented online monitoring and data quality
assurance utilizing time-stamped QA histograms produced by the ALICE High Level Trigger. The system
receives the data via ZeroMQ, stores it for later review, enriches it with detector specific functionality,
and visualizes it via a web application. These provided capabilities are complementary to the existing Data
Quality Monitoring system. In addition to basic visualization, additional processing options are
accessible to the user, including requests for data within a particular time range or reprocessing of a
particular run with different processing parameters. For example, the first ten minutes of a particular run
could be investigated for a transient hot trigger channel.

Due to similarities between the Overwatch architecture and that which will be used for Quality Control (QC)
in LHC Run 3 and beyond, Overwatch will also be utilized to develop and test various QC components during LHC Run 2.
Some of the areas of QC development include a new trending and alarm framework. We report on the project's design,
development, and status.
}
\maketitle
\hypertarget{sect:intro}{\section{Introduction}\label{sect:intro}}

ALICE (A Large Ion Collider Experiment) \cite{alice} is the dedicated
heavy-ion experiment at the Large Hadron Collider (LHC), with the goal
of studying the properties of the deconfined state of matter known as
the Quark-Gluon Plasma. Given the high particle-multiplicity physics
environment, large data rates, and complexity of the detectors,
effective data-quality monitoring is an integral part of recording
useful physics data. For the ALICE collaboration, this important task
has been successfully addressed during LHC Run 1 and 2 by the
data-quality monitoring (DQM) system known as AMORE \cite{aliceDQM}.

Despite the success of AMORE, there are still remaining open questions
related to data-quality monitoring within ALICE, particularly with
regard to the increased data rates in LHC Run 3 and beyond due to the
continuous readout of the ALICE apparatus. In order to address these
questions, we can leverage the ALICE High Level Trigger (HLT)
\cite{HLTWhitePaper,HLTRun2Performance}, which provides substantial
additional processing power beyond that available in AMORE. There are
two main questions that must be investigated in order to determine the
usefulness of an HLT-based online monitoring system. First, what novel
data-quality monitoring capabilities could be implemented during Run 2
by taking advantage of this additional processing power? Second, looking
further forward, ALICE is planning major upgrades for Run 3, which
include a substantial software project known as \osquared{} \cite{o2TDR}
-- since the current HLT operates in a similar manner as \osquared{},
what experiences related to operating a quality assurance (QA) system in
such an environment could be developed while still in Run 2?

In an effort to address these questions, the ALICE Overwatch
project\footnote{Online Visualization of Emerging tRends and Web
  Accessible deTector Conditions using the HLT.}
\cite{overwatchSoftware} was started in November 2015 to monitor and
visualize the QA information provided by the ALICE HLT. The project
handles all steps from receiving the initial data to displaying enhanced
visualizations of processed data through a web application. To further
enhance the utility of this project for different users, a plugin system
allows for ALICE detectors to customize processing, trending, and
visualizing the data.

In comparison to the existing DQM system, Overwatch provides
complementary functionality, with a focus on providing information most
useful to expert level users. By taking advantage of the preprocessed
and time-stamped nature of the HLT data, Overwatch is able to provide
certain unique capabilities within ALICE. In particular, the data
received by Overwatch has already been highly preprocessed by the HLT,
such that the data rate is substantially reduced compared to the raw
ALICE data rate. Overwatch stores \(\sim350+\) GB/year, with the volume
increasing with each year of data taking. Given this manageable volume
of data, Overwatch is able to store it persistently. The time-stamping
of the data enables unique exploration of the evolution of data quality
within a particular run, fill, or longer time period. Moreover, such
time-series data could be useful for training machine learning models
for automated QA monitoring.
 \hypertarget{sect:overwatchDetails}{\section{ALICE Overwatch}\label{sect:overwatchDetails}}

\begin{figure}[th]
    \centering
    \includegraphics[width=10cm]{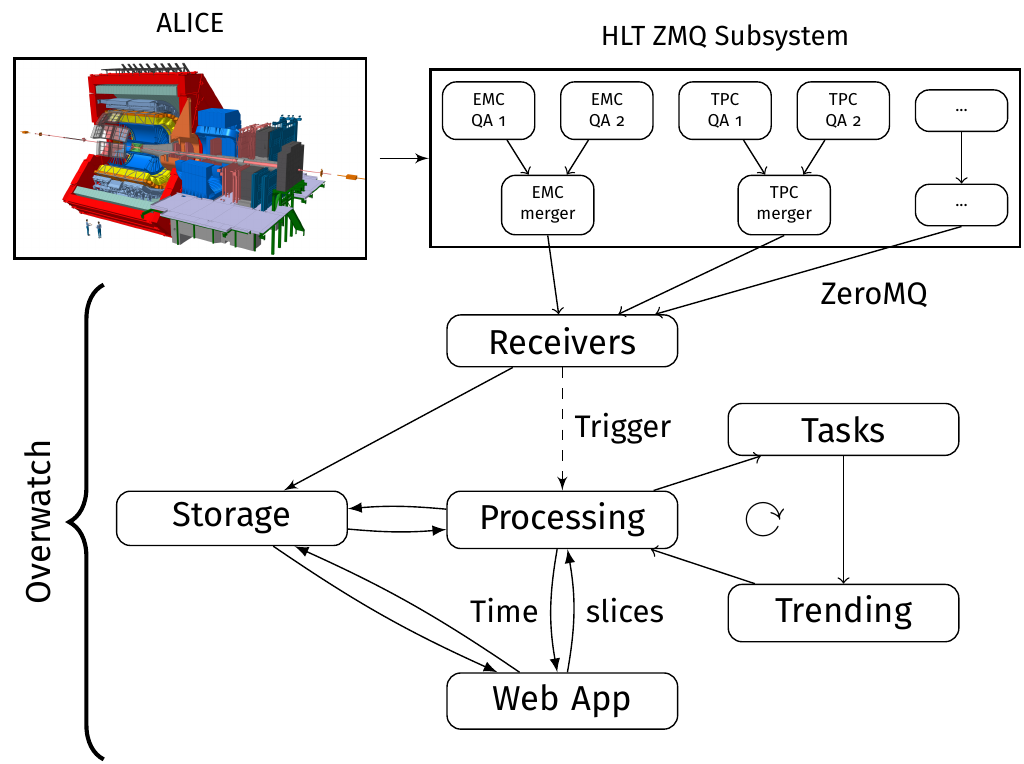}
    \caption{The Overwatch architecture, illustrating the data flow from collection in the ALICE apparatus through visualization in the Overwatch web application. Data flows along the arrows in the diagram. The HLT ZMQ subsystem, which shows contributions from the ALICE EMC and TPC detectors, represents only a small portion of the High Level Trigger. The processing tasks and trending are performed for each input object (such as a histogram) within the received data, as represented by the circular arrow.}
    \label{fig:overwatchArchitecture}
\end{figure}

The Overwatch architecture illustrating the workflow from data collected
in ALICE to final visualization in the Overwatch web application is
shown in \figRef{fig:overwatchArchitecture}. The entire process starts
when particles are detected and recorded by the ALICE apparatus. Every
recorded event is sent to the HLT. Within the HLT, the data are
reconstructed and compressed, and the output is stored, as well as
provided for further processing and analysis. Of particular significance
to Overwatch are the detector specific QA components which are
distributed throughout the HLT and run in parallel to process the
reconstructed data and extract information relevant for each detector.
Each distributed component is provided a subset of the reconstructed
data, and the output is sent to a group of centralized mergers per
detector via the ZeroMQ (ZMQ) messaging library \cite{zeroMQSoftware}.
The QA component outputs are merged and are made available for further
processing via ZMQ. The model for data flow from the QA components
through the output of the data mergers is compatible with that of
\osquared{}.

The Overwatch receiver component is written in C++ \cite{c++11Software}
and makes requests for data to the HLT data mergers approximately every
minute. However, the round-robin nature of how the data are distributed
to the QA components means that the time resolution is on the order of a
few minutes, with a strong dependence on the properties and processing
rate of a particular component \cite{HLTWhitePaper}. The messages
provided via ZMQ consist of metadata, as well as collections of
ROOT-based objects. \texttt{TH1}-derived histograms are the primary type
of data received, although other more complicated objects, such as
\texttt{TObjArray}, are also utilized. Each request contains cumulative
data, such that the difference between the objects in the currently
received message and the previous message corresponds to the data that
was collected in the time between the requests. Objects are only reset
at the beginning of each run, which typically last on the order of
several hours. In 2015, the Electromagnetic Calorimeter (EMCal)
implemented a QA component, while the Time Projection Chamber (TPC)
added a QA component in 2016. The HLT itself also provides some QA
information generated during the data reconstruction. As an example of
the information provided by a detector, during the 2015 Pb--Pb data
taking period, the EMCal QA component provided tower energy spectra,
number of triggers vs.~position, and median vs.~maximum tower energy to
characterize the event background, as well as a number of other
histograms. All of these were provided as a function of different
regions in the detector. It is worth noting that the data volume depends
heavily on the data taking period and the detectors themselves, with
some payloads as small as \(\sim10\) kB/request, while others are larger
than \(1\) MB/request.

A number of processing actions are performed upon the newly received
data. These actions are managed by the Overwatch processing module,
which is written in Python \cite{pythonSoftware}, utilizes PyROOT
\cite{ROOT,ROOTSoftware}, and is generally responsible for transforming
the data into more easily comprehensible forms. The received data is
organized and metadata related to the run and the available subsystems
(detectors) are extracted. The histograms and other objects available
for each subsystem are also classified and sorted into related groups
for visualization purposes. Overwatch then enters into the main
processing routine which adds additional context, extracts derived
information, and stores enhanced visualizations (such as highlighting
trigger regions within a detector to aid in identifying channels which
fire too frequently). This step also has the capability to extract
derived values as the data are processed for trending purposes.
Afterwards, the trended information is processed in a similar manner to
the originally received data. The output of the processing and trending
includes metadata related to the runs and included subsystems, as well
as JSON and image representations of the processed data.

This information is then made available through the web application,
where the data are organized and displayed in the main view according to
run and subsystem. The web application back end is written in Python and
is built with the Flask microframework \cite{flaskSoftware}. The front
end is built on top of Google's Polymer library \cite{polymerSoftware}
and JSRoot \cite{jsRoot,jsRootSoftware}. Within a selected run and
subsystem, the data are displayed according to their previously
classified groups, along with any information extracted from the
displayed object. Predefined data exploration capabilities are provided
for the displayed data, as will be illustrated in
\secRef{sect:timeSlices}. To further enable data exploration beyond the
capabilities provided by Overwatch, the underlying unprocessed ROOT
files are also made accessible to the user.

\begin{figure}[th]
    \centering
    \includegraphics[width=11.5cm]{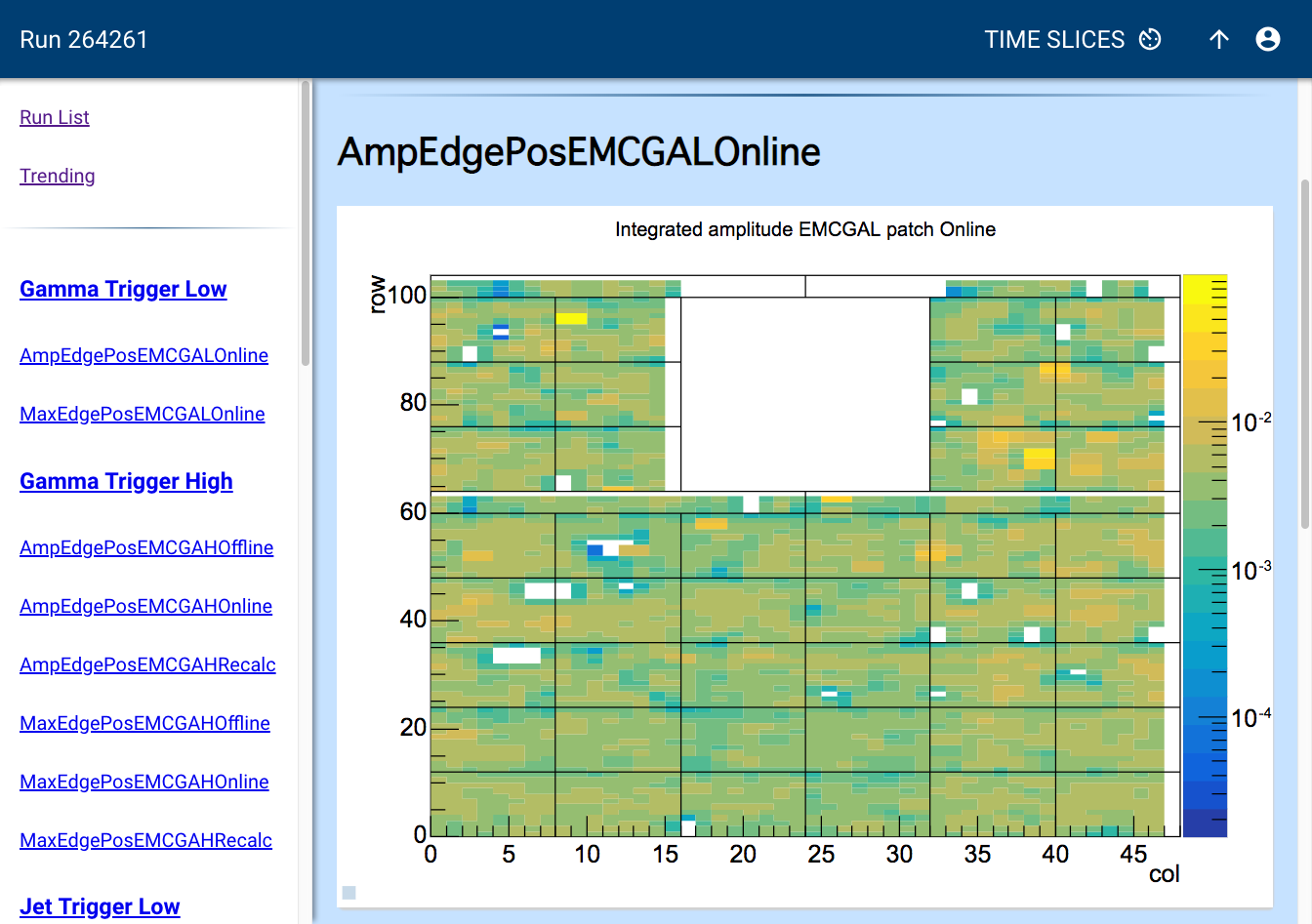}
    \caption{Typical view of the Overwatch interface showing number of triggers vs. position for the ALICE Electromagnetic Calorimeter for a particular run. The black lines superimposed on the image display trigger regions within the calorimeter to aid in the identification of hot trigger regions. See the text for further details.}
    \label{fig:overwatchInterface}
\end{figure}

A typical example of the Overwatch web application interface is shown in
\figRef{fig:overwatchInterface}. The main object is a histogram which
represents the number of triggers vs.~position in the EMCal integrated
over the length of an entire run. The white gaps which are surrounded by
data are caused by disabled trigger channels. Overwatch enhances this
visualization by superimposing a black grid which corresponds to trigger
regions within the calorimeter. Such additional content improves the
user experience by allowing identification of whether a particular
trigger region is firing too frequently (ie. ``hot'') at a glance. Since
this visualization of the data utilizes JSRoot, further investigation of
each trigger is possible by zooming in. Additional EMCal histograms are
available via the list on the left.

While there are some standard capabilities that will be used by all
detectors and therefore can be implemented by the main Overwatch
packages, each detector has a unique set of requirements. Consequently,
Overwatch implements a plugin system to allow each detector to customize
the processing, trending, and visualization of their data to meet their
needs. This is implemented through the dynamic loading of detector
specific Python modules. Each function call is then redirected to the
module if the relevant step is implemented. This approach allows each
detector to add any arbitrary functionality that is required. An example
result of this system can be seen in \figRef{fig:overwatchInterface},
where the black grid was superimposed on the displayed data via a
detector plugin in the processing module.
 \hypertarget{sect:timeSlices}{\subsection{Time slices and reprocessing}\label{sect:timeSlices}}

Beyond the basic functionality described above, the most important
Overwatch features revolve around taking advantage of the time-stamped
persistently stored data. To utilize the time-stamps, Overwatch allows
the user to request that data within a user specified start and end time
be reprocessed, thereby only using information within that time window.
This is a major data exploration feature in Overwatch, enabling
characterization of the time dependence of the data quality. This
capability is known as time slices.

\begin{figure}[th]
    \begin{center}
        \includegraphics[width=11.5cm]{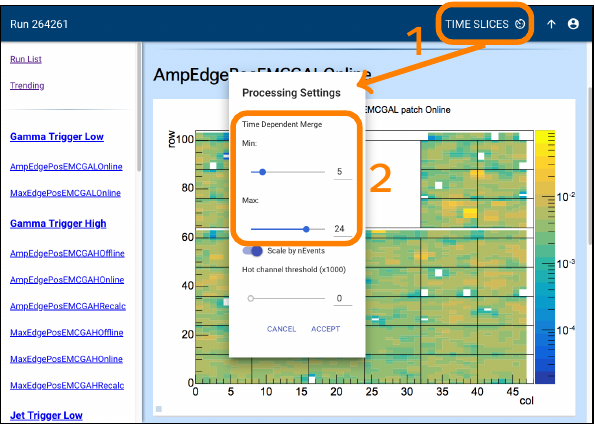}
    \end{center}
    \caption{The Overwatch time slices interface allows for the reprocessing of data within a specified time window. To perform a time slice, the user opens the time slice panel by clicking on the (1) highlighted "time slice" button. This opens a panel that is shown in the middle of the image. Within this panel, the user selects a time window within the (2) highlighted region. Five minutes from the start of the run to 24 minutes into the run is selected. Once submitted, the resulting histogram would be displayed in the same page.}
    \label{fig:overwatchTimeSlices}
\end{figure}

Overwatch exposes this functionality directly to the user through the
web application interface shown in \figRef{fig:overwatchTimeSlices}. To
create a new time slice, the user selects the highlighted ``time
slices'' button, which opens a new panel. This panel allows the user to
specify the minimum and maximum times of their time slice, as shown in
the second highlighted region. After processing the data within the
specified time window, the results are then displayed in the same page.
These results are stored and cached based on the input parameters to
allow for immediate results and an improved user experience if the time
slice has already been processed.

To illustrate the value of this approach, consider the case of a
histogram which measures the occurrence of large amplitude hits in a
detector vs.~their position. Even if the histogram indicates that a
channel had some large amplitude entries, the user doesn't necessarily
know the cause. However, if they have some external information, such as
the knowledge that the electronics initialization at the start of run
can cause large amplitude spikes that have no impact on the data quality
of the run, time slices allow the user to directly check whether these
large amplitude entries are problematic. The user would simply deselect
the first few minutes of the run, and if the large amplitude counts
disappear, then the data quality for that run is still good.

The time slice reprocessing is performed by the exact same code as the
standard Overwatch processing, including all detector plugins -- it just
utilizes different input data corresponding to the selected time window.
This procedure can then be generalized further to allow the user to
specify any exposed set of processing parameters. Such an approach adds
a new dimension to the data exploration capabilities provided by
Overwatch, enabling the extraction of custom information, as well as the
selection of different areas of the parameter space which might be
otherwise inaccessible behind other data.

\begin{figure}[th]
    \begin{center}
        \includegraphics[width=11.5cm]{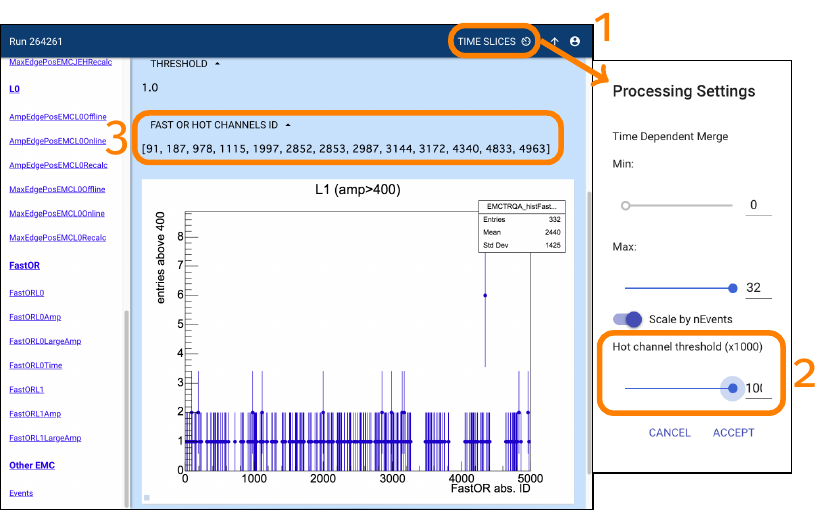}
    \end{center}
    \caption{Overwatch allows the reprocessing of histograms with different parameters. An EMCal histogram is shown which measures large amplitude hits vs. channel identifier. To perform the reprocessing, the user opens the reprocessing panel (1) by clicking on the highlighted "time slices" button. This opens a panel that is shown on the right. In this example, the user selects a threshold (which is multiplied by a trivial scaling factor for display purposes) in (2), which specifies the maximum counts threshold for the reprocessing. The text (3) displays the channels found above this threshold, with the reprocessed histogram displayed below. Note that in this example the reprocessed histogram is identical to the original histogram, but in general this need not be the case.}
    \label{fig:overwatchReprocessing}
\end{figure}

As an example, consider again a histogram which measures the occurrence
of large amplitude hits in a detector vs.~their position. If there has
been a particularly noisy run, the user can extract all channels (bins)
whose entries are above a user specified threshold in order to take
further actions. \figureRef{fig:overwatchReprocessing} illustrates just
such a histogram from the EMCal and demonstrates how these reprocessing
capabilities can be used. The user first selects the reprocessing and
time slices button, which is labeled with ``time slices'' and is
highlighted in the first selection. This opens a panel to direct the
reprocessing. The user can select their desired threshold for maximum
number of counts, as shown in the second highlighted selection. Once
submitted, the data will be reprocessed, and the identifiers of the hot
channels are displayed in a list inside of the third highlight,
alongside the original histogram. With this information, a detector
expert can take the appropriate actions to address the identified
channels. As with all other parts of Overwatch, these capabilities are
extensible, such that new reprocessing options can be added by any
detector subsystems. Options which are not relevant for the currently
displayed histogram will be disabled. Note that it is also acceptable to
specify a time slice window along with other reprocessing options.
 \hypertarget{sect:outlook}{\section{Outlook towards Run 3}\label{sect:outlook}}

While most of the Overwatch infrastructure will shut down at the end of
2018 with the completion of LHC Run 2, it can still provide substantial
contributions before the run is completed, as well as afterwards. As
noted in \secRef{sect:overwatchDetails}, the data model for the HLT QA
data is compatible with that of \osquared{}. Furthermore, the Overwatch
architecture to handle and process data closely matches the design of
the ALICE Quality Control (QC) project for data-quality monitoring in
\osquared{}. Although Overwatch processes much less data than will be
handled during Run 3, the strong similarities demonstrate that Overwatch
can inform future developments. Consequently, in addition to maintaining
the project's existing capabilities, Overwatch is being utilized to
develop experiences and prototypes necessary for \osquared{} in Run 3
while still in Run 2. The focus of these efforts is on the trending and
alarms subsystems in Overwatch. The existing systems have fairly basic
capabilities, so new developments are ongoing which attempt to satisfy
the requirements of both Run 2 and Run 3. The practical knowledge, and
perhaps the code itself, will help inform the development of the
corresponding systems within the QC project.

Post Run 2, Overwatch can continue to provide valuable experience by
taking advantage of the persistently stored data. This data can be
replayed as if it was just received from the HLT. For any QC related
developments, prototypes can be quickly implemented through the
Overwatch plugin system, allowing testing of new ideas with real data.
The time-stamped nature of the data can also be useful as time-series
data for training machine learning models. Such capabilities ensure that
Overwatch will continue to make contributions beyond 2018.
 \hypertarget{sect:conclusions}{\section{Conclusions}\label{sect:conclusions}}

ALICE Overwatch is a project developed in Python and C++ to process and
visualize quality assurance data that is produced by taking advantage of
the additional processing power of the ALICE High Level Trigger.
Overwatch uses this persistently stored, time-stamped data, to provide
unique data-quality monitoring capabilities within ALICE. A plugin
system is provided for detectors to customize the processing, trending,
and visualization of the stored TB-sized Run 2 dataset. The output of
this processing is displayed in a web application which facilitates data
exploration through reprocessing based on user specified parameters, as
well as within time slices. Due to the similarity between the
architecture of Overwatch and the Quality Control project for ALICE
\osquared{}, the capabilities and experiences described here can be used
to inform the development of the QC system, with a particular focus on
the trending and alarms subsystems.
 
\section*{Acknowledgments}

We thank Markus Fasel, Mikolaj Krzewicki, Mateusz Ploskon, and David Rohr for useful discussions, advice, and help in starting the project that became Overwatch. We also thank the ALICE QA Tools and WP7 working groups for useful discussions and advice. This work was supported in part by the U.S. Department of Energy, Office of Science, Office of Nuclear Physics under Grant number DE-SC004168.

\bibliography{chepProceedings}

\begin{thebibliography}{15}

\bibitem{alice}
{ALICE Collaboration}, Journal of Instrumentation \textbf{3}, S08002 (2008)

\bibitem{aliceDQM}
B.~von Haller, A.~Telesca, S.~Chapeland, F.~Carena, W.~Carena, V.C. Barroso,
  F.~Costa, E.~Denes, R.~Divi{\`a}, U.~Fuchs et~al., Journal of Physics:
  Conference Series \textbf{331}, 022030 (2011)

\bibitem{HLTWhitePaper}
{ALICE Collaboration}, In preparation  (2018)

\bibitem{HLTRun2Performance}
M.~Krzewicki, V.~Lindenstruth, {for the ALICE Collaboration}, Journal of
  Physics: Conference Series \textbf{898}, 032056 (2017)

\bibitem{o2TDR}
P.~Buncic, M.~Krzewicki, P.~Vande~Vyvre, Tech. Rep. CERN-LHCC-2015-006.
  ALICE-TDR-019 (2015), \urlstyle{tt}\url{https://cds.cern.ch/record/2011297}

\bibitem{overwatchSoftware}
R.~Ehlers, J.~Mulligan, \emph{{ALICE Overwatch} [software] v1.0} (2018),
  \urlstyle{tt}\url{https://doi.org/10.5281/zenodo.1309376}

\bibitem{zeroMQSoftware}
{ZeroMQ Project}, \emph{Zeromq [software, v 4.2.5} (2018),
  \urlstyle{tt}\url{http://zeromq.org/intro:get-the-software}

\bibitem{c++11Software}
\emph{{ISO\slash IEC 14882:2011 Information technology --- Programming
  languages --- C++}} (International Organization for Standardization, 2011)

\bibitem{pythonSoftware}
{Python project}, \emph{Python [software], v. 3.6.6} (2018),
  \urlstyle{tt}\url{https://www.python.org/downloads/release/python-366/}

\bibitem{ROOT}
R.~Brun, F.~Rademakers, Nucl. Instrum. Meth. \textbf{A389}, 81 (1997)

\bibitem{ROOTSoftware}
{ROOT project}, \emph{Root [software], v 6.10/08} (2018),
  \urlstyle{tt}\url{https://root.cern.ch/content/release-61008}

\bibitem{flaskSoftware}
{Pallets Project}, \emph{Flask [software], v 1.0.2} (2018),
  \urlstyle{tt}\url{https://github.com/pallets/flask/releases/tag/1.0.2}

\bibitem{polymerSoftware}
{Polymer Project}, \emph{Polymer [software] v1.0} (2018),
  \urlstyle{tt}\url{https://www.polymer-project.org/1.0/start/}

\bibitem{jsRoot}
B.~Bellenot, S.~Linev, Journal of Physics: Conference Series \textbf{664},
  062033 (2015)

\bibitem{jsRootSoftware}
B.~Bellenot, S.~Linev, \emph{{JavaScript ROOT} [software], v. 5.5.0} (2018),
  \urlstyle{tt}\url{https://github.com/root-project/jsroot/releases/tag/5.5.0}

\end{thebibliography}

\end{document}